\documentclass[12pt,twoside]{article}
\usepackage[super,sort,comma]{natbib}
\usepackage{fancyhdr}

\usepackage{showkeys}		

\usepackage[section]{placeins}
\usepackage{graphicx}
\usepackage{subfigure}
\usepackage{parskip}
\usepackage{pdfpages}
\usepackage{tabularx}
\usepackage{threeparttable}
\usepackage{amssymb}
\usepackage{epstopdf}
\usepackage{plain}
\usepackage{amsmath}
\usepackage{multirow}
\usepackage[bottom]{footmisc}

\makeatletter \renewcommand\@biblabel[1]{$^{#1}$} \makeatother
\usepackage[mathlines]{lineno}
\usepackage{hyperref}
\hypersetup{ colorlinks,
    citecolor=blue,
    filecolor=blue,
    linkcolor=blue,
    urlcolor=blue
}
\usepackage{xcolor}
\definecolor{gray}{rgb}{0.6,0.6,0.6}
\definecolor{red}{rgb}{0.85,0,0}
\definecolor{green}{rgb}{0,0.85,0}
\definecolor{blue}{rgb}{0,0,0.85}
\definecolor{beige}{rgb}{0.92,0.87,0.78}
\usepackage[all]{hypcap}    %causes link to figures to go to figure, not caption
\begin{document}

% \cen{
% \sf 
{\Large {\bfseries DECT-based Space-Squeeze Method for Multi-Class Classification of Metastatic Lymph Nodes in Breast Cancer } \\  
\vspace*{10mm}
Hai Jiang$^{\dagger, 1}$, Chushan Zheng$^{\dagger, 3}$, Jiawei Pan$^1$, Yuanpin Zhou$^4$, Qiongting Liu$^5$, Xiang Zhang$^{*, 2}$, Jun Shen$^{*, 2}$, Yao Lu$^{*, 1}$ \\
$^1$School of Computer Science and Engineering, Sun Yat-sen University, P.R. China \\ 
$^2$Department of Radiology, Sun Yat-Sen Memorial Hospital, Sun Yat-Sen University, P.R. China \\ 
$^3$Medical Imaging Center, Shenzhen Hospital of Southern Medical University, P.R. China \\
$^4$Hithink RoyalFlush Information Network Co., Ltd., Hangzhou, P.R. China \\
$^5$Department of Radiology, The third Affilliated Hospital of Guangzhou Medical University, P.R. China \\
\vspace{5mm}\\
Version typeset \today\\
% }

\pagenumbering{roman}
\setcounter{page}{1}
\pagestyle{plain}
$^{\dagger}$ Equal contribution\\
$^*$ Correspondence\\
email: luyao23@mail.sysu.edu.cn, shenjun@mail.sysu.edu.cn, zhangx345@mail.sysu.edu.cn \\

\begin{abstract}
\noindent {\bf Background:}
Accurate assessment of metastatic burden in axillary lymph nodes is crucial for guiding breast cancer treatment decisions, 
yet conventional imaging modalities struggle to differentiate metastatic burden levels and 
capture comprehensive lymph node characteristics. 
This study leverages dual-energy computed tomography (DECT) to exploit spectral-spatial information for 
improved multi-class classification.\\ 
{\bf Purpose:}
To develop a noninvasive DECT-based model classifying sentinel lymph nodes into three categories: 
no metastasis ($N_0$), low metastatic burden ($N_{+(1-2)}$), and heavy metastatic burden ($N_{+(\geq3)}$), 
thereby aiding therapeutic planning.\\
{\bf Methods:}
We propose a novel space-squeeze method combining two innovations: 
(1) a channel-wise attention mechanism to compress and recalibrate spectral-spatial features across 11 energy levels, 
and (2) virtual class injection to sharpen inter-class boundaries and compact intra-class variations 
in the representation space.\\
{\bf Results:}
Evaluated on 227 biopsy-confirmed cases, 
our method achieved an average test AUC of 0.86 (95\% CI: 0.80-0.91) across three cross-validation folds, 
outperforming established CNNs (VGG, ResNet, etc). 
The channel-wise attention and virtual class components individually improved AUC by 5.01\% and 5.87\%, respectively, 
demonstrating complementary benefits.\\
{\bf Conclusions:}
The proposed framework enhances diagnostic AUC by effectively 
integrating DECT's spectral-spatial data and mitigating class ambiguity, 
offering a promising tool for noninvasive metastatic burden assessment in clinical practice.\\
\end{abstract}

\newpage

\tableofcontents

\newpage

\setlength{\baselineskip}{0.7cm}

\pagenumbering{arabic}
\setcounter{page}{1}
\pagestyle{fancy}
\section{Introduction}
Breast cancer (BC) has been the most commonly diagnosed cancer worldwide from 2000 to 2022, 
surpassing Lung cancer in incidence (2.25 million BC cases vs. 2.21 million Lung cancer cases) 
\cite{sung2021global,bray2024global,lei2021global}. 
The sentinel lymph node (SLN), the first node to drain the primary tumor, 
plays a critical role in treatment planning and determining the necessity of surgical intervention, 
potentially avoiding unnecessary biopsies \cite{montemurro2012omission}. 
Based on nodal dissection pathology, SLN metastatic status is categorized as disease-free axilla ($N_0$), 
low metastatic burden ($N_{+(1-2)}$), or heavy metastatic burden ($N_{+(\geq3)}$) 
\cite{egner2010ajcc,giuliano2011axillary}. 
For SLN-positive patients, axillary lymph node dissection (ALND) 
has been the standard procedure to evaluate axillary lymph node metastasis and manage regional nodes. 
However, ALND can lead to complications such as upper limb edema, neuropathic pain, and arm numbness, 
particularly in patients with low metastatic burden \cite{savolt2017eight}. 
Emerging evidence suggests that ALND may be avoidable for SLN-positive patients with $N_{+(1-2)}$ status \cite{gradishar2022breast}. 
Therefore, accurately identifying ALN status is essential, 
serving as a key indicator for cancer staging and guiding therapeutic decisions for BC patients.
\begin{figure}[htbp]\label{fig1}
  \centering
  \begin{minipage}[b]{\linewidth}
    \subfigure[]{
      \includegraphics[width=0.98\linewidth]{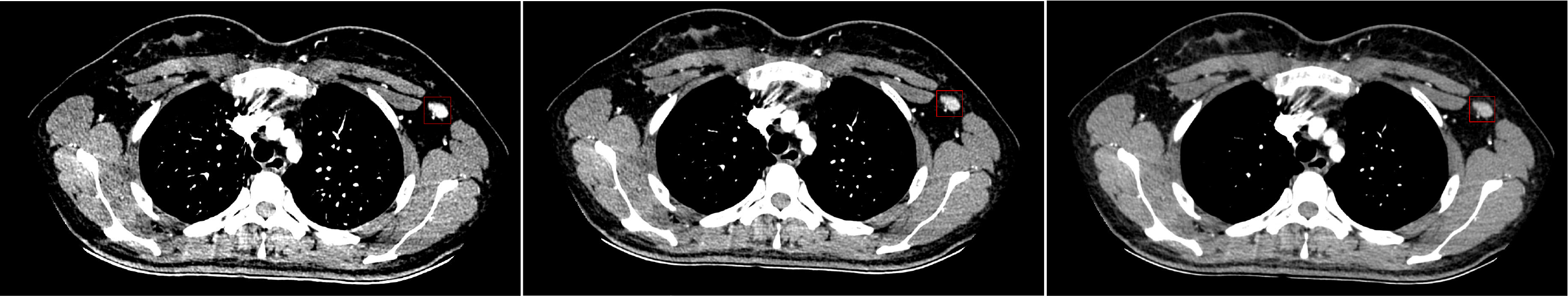}
    }
  \end{minipage}
  \centering
  \begin{minipage}{\linewidth}
    \subfigure[]{
      \includegraphics[width=0.98\linewidth]{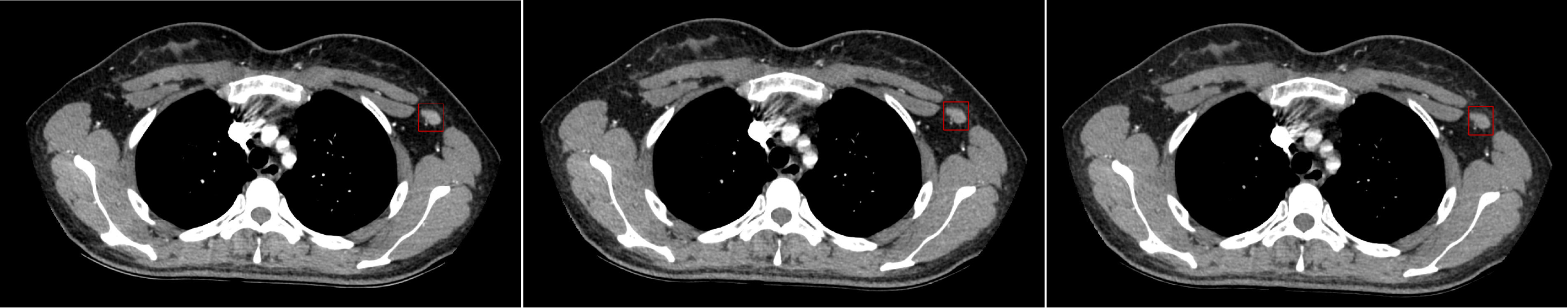}
    }
  \end{minipage}
  \centering
  \begin{minipage}{\linewidth}
    \subfigure[]{
      \includegraphics[width=0.98\linewidth]{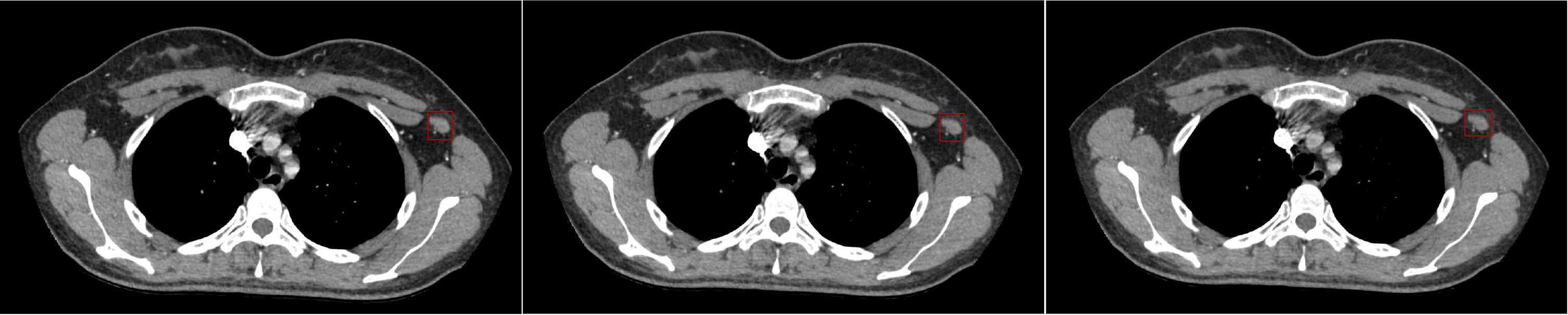}
    }
  \end{minipage}
  \centering
  \begin{minipage}{0.7\linewidth}
    \subfigure[]{
      \includegraphics[width=0.97\linewidth]{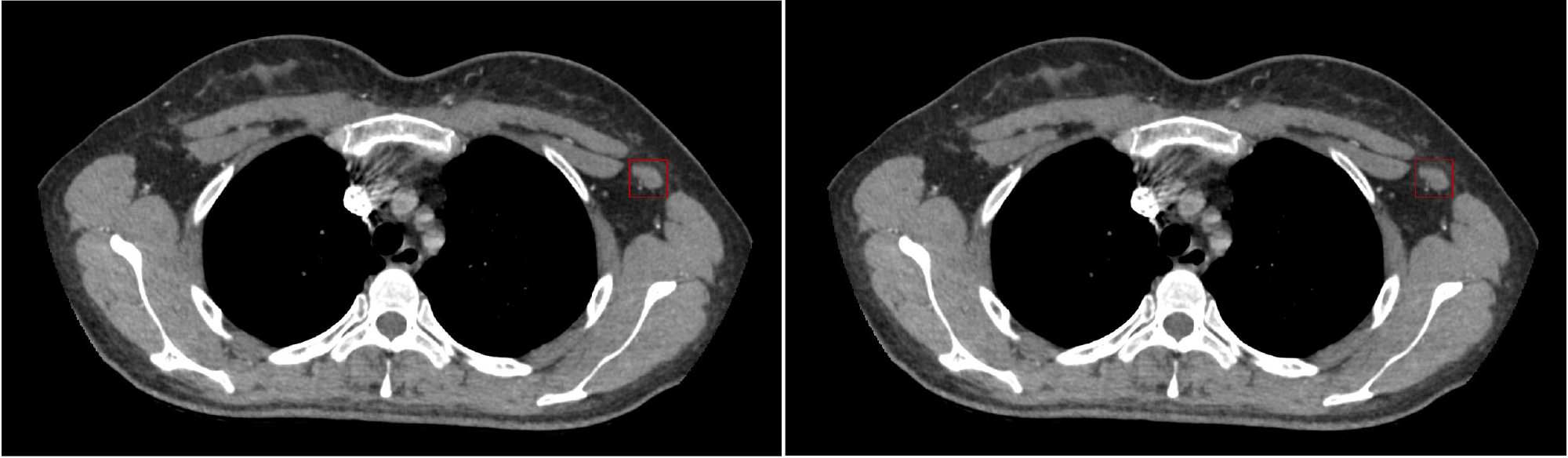}
    }
  \end{minipage}
  \caption{Subfigures (a) to (d) depict a patient with heavy-burden metastatic disease ($N_{+(\geq3)}$) 
          across 11 energy levels, ranging from 40 keV to 140 keV in 10 keV increments. 
          The sentinel axillary lymph node is marked with a red bounding box.}
\end{figure}

Ultrasound (US) and histopathology are widely used for diagnosis but offer limited accuracy and require large datasets, 
making them less effective in small-sample settings \cite{youk2017pre,2019Predicting}. 
Similarly, MRI and PET have shown suboptimal performance compared to SLN biopsy \cite{ecanow2013axillary}. 
Although CT is commonly used for staging or detecting metastases in advanced BC, 
its preoperative application, particularly in predicting SLN metastasis, 
remains limited, often due to clinician inexperience. 
In contrast, dual-energy computed tomography (DECT) is emerging as 
a promising complementary modality \cite{zhang2018axillary}. 
DECT combines spectral Hounsfield unit curves with spatial imaging, 
allowing for the assessment of both tissue composition and lymph node metastasis. 
Its ability to capture tissue absorption differences across energy levels and provide high-resolution images  
enables a more comprehensive and detailed evaluation of lymph node characteristics in BC patients.

Existing methods mainly focus on detecting the presence or absence of sentinel ALN metastasis, 
typically classifying cases as $N_0$ and non-$N_0$. 
For instance, Zhou et al. employed two deep learning models 
to predict ALN metastasis in BC patients \cite{zhou2020lymph}, 
highlighting the potential of deep learning in this domain. 
Other studies using US data have applied statistical machine learning models for metastasis diagnosis 
\cite{evans2014does,boughey2010cost,qiu2016nomogram}. 
La Verde et al. used logistic regression to examine tumor and patient characteristics influencing 
SLN involvement in early-stage BC \cite{la2016role}, 
while Tapia et al. applied logistic regression to pathological images, 
achieving an AUC of $0.74$ \cite{tapia2019predicting}. 
In contrast, our study moves beyond binary classification by proposing a non-invasive, 
multi-class classification framework designed 
to support radiologists in making more accurate and comprehensive diagnoses.

Several challenges need to be addressed in this study: 
\begin{itemize}
  \item Frist, the use of a region of interest (RoI) that focuses on the largest sentinel ALN, 
  combined with the limited dataset, increases the risk of overfitting in the deep CNN methods.
  \item Second, DECT's unique properties, 
  which combine spectral information (Hounsfield unit curves) with spatial data across multiple energy levels, 
  present the challenge of effectively extracting and utilizing both types of information.
  \item Third, sentinel ALNs exhibit variations across different energy levels, 
  making it difficult for radiologists to discern discrepancies (Fig. \ref{fig1}) in clinical scenarios. 
  Capturing and leveraging the correlated information across these energy levels is therefore a significant challenge. 
  \item Finally, the morphological similarities across different classes (Fig. \ref{fig2}) 
  increase the difficulty of classification, 
  as the hidden layers in CNNs tend to generate overlapping feature patterns 
  that are challenging to control during training. 
  Effectively managing these representations remains a critical challenge.
\end{itemize}

In this work, we propose a novel space-squeeze approach to address these issues. 
The manuscript is organized as follows: 
Section II introduces the materials, the proposed diagnostic model, and experimental setup. 
Section III presents the experimental results. 
Section IV provides an in-depth discussion of the findings. 
Finally, Section V gives the conclusions.

\section{Materials and Methods}
\subsection{Data}\label{dataset}
Prior to this retrospective study, ethical approval was obtained from 
the Institutional Review Board at Sun Yat-sen Memorial Hospital, 
and written informed consent was secured from all patients. 
A total of 458 patients with histopathologically confirmed BC were initially recruited. 
All underwent chest and axillary spectral CT scans using a GE Discovery CT750HD scanner, 
with patients positioned supine. 
Dual-phasic contrast-enhanced imaging was performed using Iohexol (GE Healthcare, USA), 
injected at 4 $ml/s$ via the ulnar vein with an automated injector (Medrad Stellant). 
Arterial-phase imaging employed semi-automatic contrast tracking. 
Of the 458 patients, 227 were included in the final analysis based on confirmed metastatic status. 
The remaining 231 were excluded for reasons such as prior therapy (47), pre-scan biopsy (105), 
benign lesions (74), micrometastasis (1), missing intraoperative biopsy (2), and data integrity issues (2). 
Among the included patients, 96 had confirmed metastasis (58 low burden, 38 heavy burden), 
while 131 were metastasis-free.

To evaluate model performance, the 227-patient dataset was split into training (151 cases) 
and test (76 cases) sets with stratification by class labels. 
Training utilized 3-fold stratified cross-validation. 
Image pixel values were normalized using a background-aware formula where $-2000$ 
denoted background intensity and $p^+_{min}$ was the minimum non-background pixel value in the training fold. 
Model performance was assessed using the area under the receiver operating characteristic curve (AUC). 
Each patient's data included 11 energy levels (40-140 keV in 10 keV intervals). 
An experienced radiologist (Xiang Zhang, 10 years in BC diagnosis, 5 years in segmentation) 
manually annotated the region of interest (RoI) along the largest lymph node contour on the largest axial slice. 
RoIs were cropped, resized to $176\times176$ pixels via bilinear interpolation, and centered for consistency.

\subsection{Architecture of the CNN-Based Classification Pipeline}
Building on the strengths of CNNs, our proposed method emphasizes efficiency 
through a shallow network architecture to minimize the risk of overfitting. 
As illustrated in Fig. \ref{overview}, 
the model builds upon our previous work \cite{zeng2021decoupling} 
and features four convolutional blocks (highlighted in yellow). 
The first block processes multi-channel input using a $1\times1$ kernel with a stride of $1$, 
while the second and third blocks use $3\times3$ kernels with stride $2$ 
to extract spectral and spatial features, respectively. 
This architecture, consisting of an embedding layer followed by three convolutional layers, 
is designed to decouple and integrate spectral and spatial information—effectively addressing the second challenge.

A feature-space squeezing module is integrated at the bottom-right of the first convolutional block, 
enhancing the model's ability to refine learned features. 
Additionally, virtual class injection begins at the model's input, 
with the corresponding injected fake label (depicted in red) incorporated into the model's output. 
\begin{figure}[htbp]
  \centering
  \includegraphics[width=0.98\textwidth]{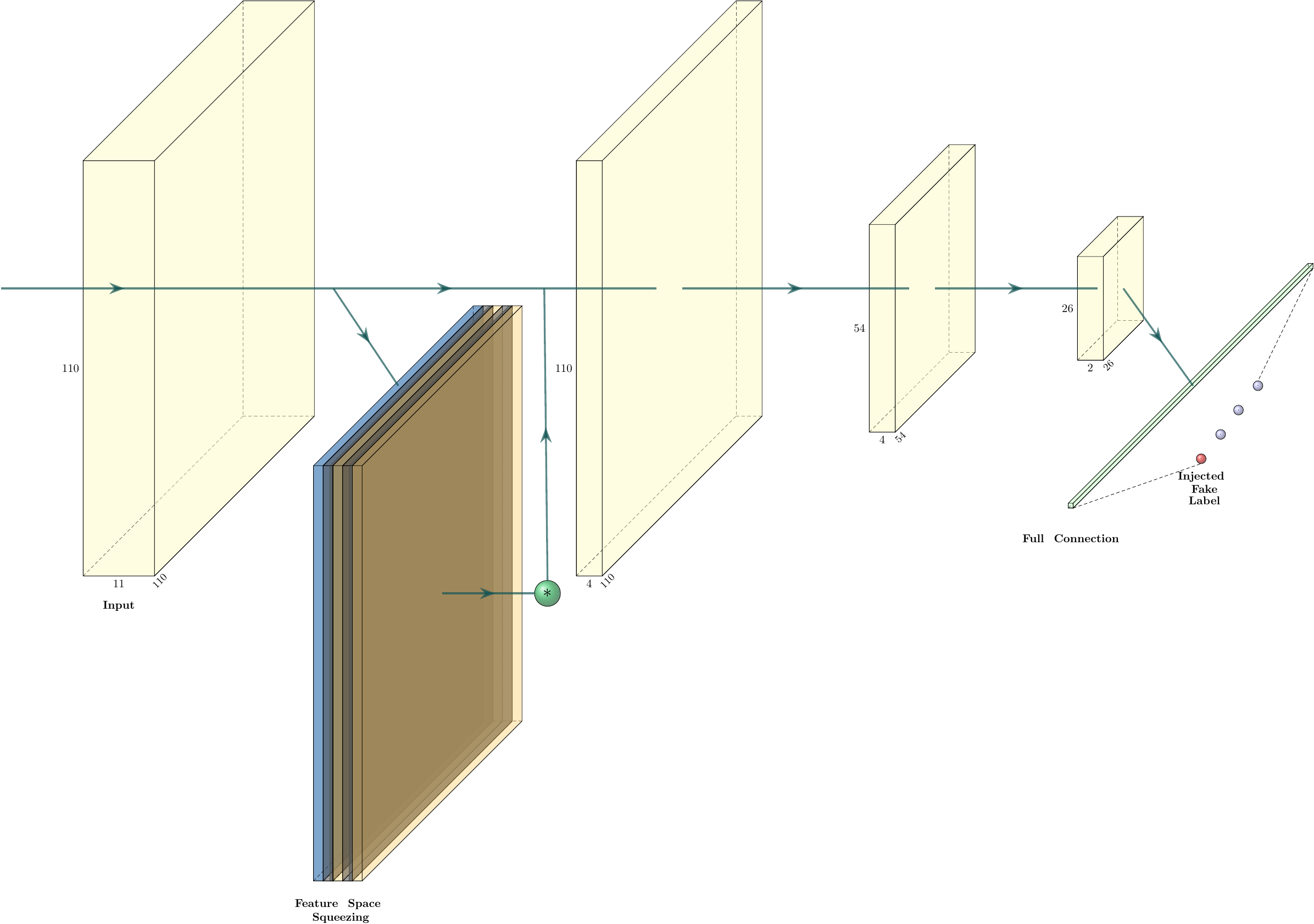}\label{overview}
  \caption
  {The proposed CNN-based framework.}
\end{figure}

\subsubsection{Feature space squeezing}\label{fss}
To leverage the correlated information across the 11 selected energy levels, 
we compressed the feature map in the space encoded from images at all levels, 
and then re-excited it using learnable coefficients from an additional shallow embedding layer. 
By squeezing the feature space, our model aims to enhance its representational power, 
as the re-excitation process recalibrates features from different energy levels, 
explicitly modeling the interdependencies between them. 
\begin{figure}[htbp]
  \centering
  \includegraphics[width=0.98\textwidth]{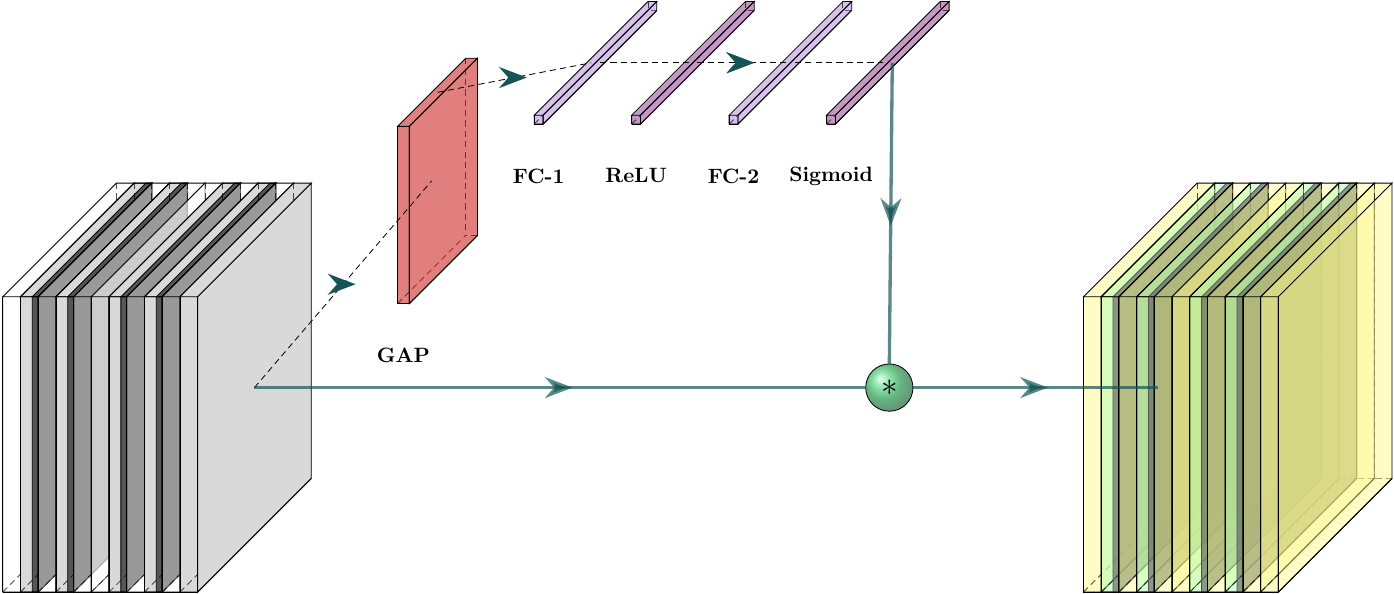}\label{seblock}
  \caption{The detailed framework of feature-space-squeezing module. }
\end{figure}

We define the feature map in the feature space as $\mathbf{X}\in\mathbb{R}^{H\times W\times C}$, 
a 4D tensor where $H$, $W$ and $C$ denote height, width, and number of channels, respectively. 
The nonlinear mapping function $T$, implemented via a \textbf{Squeeze-and-Excitation} (SE) block, 
transforms $\mathbf{X}$ into $\mathbf{X'}\in\mathbb{R}^{H'\times W'\times C'}$ 
as $\mathbf{X'}=\mathcal{T}(\mathbf{X})=\mathbf{X}\circledast\overrightarrow{\omega}$, 
where $\overrightarrow{\omega}\in\mathbb{R}^{C\times\frac{C}{r}}$ is a learnable coefficient vector 
and $\circledast$ denotes the element-wise multiplication. 
Unlike pixel-wise averaging across energy levels, 
$\overrightarrow{\omega}$ is generated through a compact embedding consisting of a Global Average Pooling (GAP) layer, 
followed by two Fully Connected (FC) layers with ReLU and Sigmoid activations. 
This structure captures channel-wise dependencies across the 11 energy levels. 
Specifically, the GAP operation computes 
$\delta_k = \frac{1}{H\times W}\sum\limits^H_{i=1}\sum\limits^W_{j=1}x_k(i, j)$ 
for each feature map $x_k$, resulting in the vector $\overrightarrow{\delta}\in\mathbb{R}^C$. 
The excitation step then applies two FC layers with a reduction ratio $r$ (chosen such that $\frac{C}{r}=4$) 
to reduce and then restore the dimension: 
\begin{equation}
  \begin{aligned}
  %     \eqalign{
      \overrightarrow{\omega} 
      = \mathbf{Sigmoid}(\mathrm{W}_2\cdot(\mathbf{ReLU}(\mathrm{W}_1\cdot\overrightarrow{\delta}
      +\overrightarrow{b_1}))+\overrightarrow{b_2}),%} 
  \end{aligned}
\end{equation}
where $\mathrm{W}_1\in\mathbb{R}^{\frac{C}{r}\times C}$, $\mathrm{W}_{2}\in\mathbb{R}^{C\times\frac{C}{r}}$, 
and $\overrightarrow{b_i}(i=1,2)$ are biases. 
The final output $\mathbf{X'}\in\mathbb{R}^{H'\times W'\times C'}$ incorporates both spatial and spectral context, 
enhancing representation of key energy-level features for downstream analysis.

\subsubsection{Representation space squeezing}\label{rss}
The proposed classification CNN model encodes the 4-dimensional (4D) tensor 
for each patient into a high-dimensional representation space, which may contain thousands of patterns for recognition. 
By defining appropriate optimization loss functions, 
the model aims to identify the most relevant representative patterns for each class. 
However, since internal and external tumor lesions may share similar patterns across different classes, 
the boundaries between classes can become blurred, leading to confusion for the model. 
To address this, we squeeze the representation space by introducing virtual classes, 
which helps sharpen the inter-class boundaries and compact the intra-class regions (as shown in Fig. \ref{VCvisual}), 
thereby improving class discrimination. 

\begin{figure}[htbp]\label{VCvisual}
  \centering
  \subfigure[Before]{
    \includegraphics{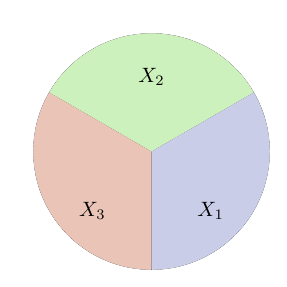}\label{bef}
  }
  \quad
  \subfigure[After]{
    \includegraphics{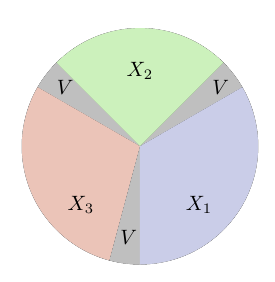}\label{aft}
  }
  \caption{A simple 2D illustration showing the representation space before Fig. \ref{bef} and after Fig. \ref{aft} 
          the injection of virtual class.}
\end{figure}
% In this subsection, 

We define the training dataset as $\{(x_i, y_i)\}$, 
where $x_i\in X$ represents the input features and $y_i\in Y =\{1, 2, ..., C\}$ denotes the ground truth label. 
The angle between output feature vectors $X_i$ and $X_{i+1}$ is $\theta_i$, 
and $||\cdot||$ denotes the $L_2$ norm. 
In standard classification, the softmax function normalizes model outputs, 
with the cross-entropy loss for class $i$ defined as $l_i=-\log\frac{\exp(W^T_{y_i}X_i)}{\sum^C_{j=1}\exp(W^T_jX_i)}$, 
where $W_j$ is the weight vector for class $j$. 
To enhance class separation, we introduce \textbf{virtual classes}, 
modifying the loss function to: 
\begin{equation}
  \begin{aligned}
    l'_i =-\log\frac{\exp(||W_{y_i}||\cdot||X_i||\cdot\cos\theta_{y_i})}
          {\sum^{C+V}_{j=1}\exp(||W_j||\cdot||X_i||\cdot\cos\theta_j)},%} 
  \end{aligned}
\end{equation}
and the overall loss becomes:
\begin{equation}
  \begin{split}
    L'=\frac{1}{C}\sum\limits^C_{i=1}l'_i.
  \end{split}
\end{equation}
This reformulation enlarges inter-class margins and compacts the decision boundaries 
by encouraging better angular separation, thereby improving classification performance.

\subsection{Experimental Configuration and Training Protocols}
Our framework was implemented on 
a platform equipped with an Intel(R) Xeon(R) CPU E5-2678 v3 @ 2.50 GHz 
and a single NVIDIA GeForce GTX 1080 Ti GPU (12 GB RAM), 
using \emph{TensorFlow} 2.4.0 as the software environment. 
During training, we applied data augmentation techniques, 
including random flipping (\emph{probability=}$0.5$) and 
random rotation within a range of $\pm18^\circ$ (\emph{probability=}$0.5$), 
using the \emph{Albumentations} library. 
The network was trained with an end-to-end approach, using a batch size of 32 and 
the \emph{Adam} optimizer with a learning rate of $1\times10^{-4}$. 
To prevent overfitting, we applied $L_2$ regularization to the weights of each Conv2D kernel 
with a regularization parameter of $1\times10^{-2}$, and used the \emph{Xavier} uniform method for kernel initialization. 
The training process spanned 500 epochs, with the training set stratified into three folds for cross-validation. 
Each input case was transformed into a $176\times176\times11$ tensor. 

Since the dataset was collected from a single-center hospital (Sun Yat-sen Memorial Hospital), 
it was divided into training and test sets as detailed in Section \ref{dataset}. 
Due to the imbalanced class distribution and limited sample size, 
we employed three-fold stratified cross-validation using the \emph{scikit-learn} library 
to ensure robust evaluation. 
Model performance was primarily measured using 
the micro-average area under the ROC curve (AUC) in a one-vs-rest (OvR) setting, 
with the highest validation AUC used for model checkpointing and final testing. 
To account for the dataset's small size and imbalance, 
the test set AUC was further analyzed using the BCa method, 
which adjusts for bias and skewness in the bootstrap distribution 
to estimate the 95\% confidence interval (CI) \cite{efron1987better}.

To ensure fair evaluation of each component on this in-house dataset, 
the random seed was fixed at 42 across all experimental libraries, 
ensuring consistent parameter initialization, data augmentation, 
and data distribution (including training, validation, and testing) 
across all methods used in our experiments. 
Instead of the focal loss used in our previous work, we employed cross-entropy loss, 
enhanced by a virtual class mechanism as part of our representation space-squeezing strategy. 
% Additional evaluation metrics included p-values for each fold, accuracy, sensitivity, specificity, and F1-score. 
To ensure comprehensive analysis, our feature-space-squeezing approach included a reduced coefficients layer, 
decreasing the number of coefficients from 11 to 4—an essential factor examined in our experiments. 
The implementation code is publicly available online 
\footnote{\url{https://github.com/pigejianghai/projects/tree/master/DECT/tfversion}}.

The 2-dimensional illustrations of the virtual class in Fig. \ref{VCvisual} 
may create confusion regarding the number of injected virtual classes. 
However, as demonstrated in the previous study \cite{chen2018virtual}, 
the virtual class strategy is primarily implemented in the label space (conceptually). 
Therefore, even if multiple injected classes are shown, 
they can be considered as a single virtual class in the context of this approach. 

\section{Results}
\subsection{Ablation Study of Key Methodological Components}\label{ablationstudy}
Building on our previous findings, 
we further evaluated the proposed method by analyzing each of its key components individually. 
In addition to the AUC scores obtained from the test set, 
we calculated the 95\% confidence intervals using the BCa method for a more robust assessment.
\begin{table}[htbp!]\label{tab1}
  \centering
  \caption{Ablation study of two components.}
  \resizebox{\linewidth}{!}{
  \begin{tabular}{c|c|c|c|c|c}
    \hline
    \multirow{2}*{SE-Block}&\multirow{2}*{Virtual-Class}&\multicolumn{3}{c|}{Test AUC - BCa 95\% CI} & Averaged\\
    \cline{3-5}
    ~ & ~ &  Fold-1 & Fold-2 & Fold-3 & AUC\\
    \hline
    \multirow{2}*{$\times$}&\multirow{2}*{$\times$}& 0.7990 & 0.7566 & 0.7400 & \multirow{2}*{0.7652}\\
    ~ & ~ & \footnotesize{$[0.7042, 0.8725]$} & \footnotesize{$[0.6565, 0.8382]$} & \footnotesize{$[0.6451, 0.8231]$}&~\\
    \cline{1-6}
    \multirow{2}*{$\surd$}&\multirow{2}*{$\times$}& 0.8093 & 0.8077 & 0.7655 & \multirow{2}*{0.8153}\\
    ~ & ~ & \footnotesize{$[0.7121, 0.8733]$} & \footnotesize{$[0.6243, 0.8067]$} & \footnotesize{$[0.6638, 0.8288]$}&~\\
    \cline{1-6}
    \multirow{2}*{$\times$}&\multirow{2}*{$\surd$}& 0.8140 & 0.8288 & 0.8289 & \multirow{2}*{0.8239}\\
    ~ & ~ & \footnotesize{$[0.7614, 0.8624]$} & \footnotesize{$[0.7675, 0.8787]$} & \footnotesize{$[0.7630, 0.8809]$}&~\\
    \cline{1-6}
    \multirow{2}*{$\surd$}&\multirow{2}*{$\surd$}& \textbf{0.8528} & \textbf{0.8861} & \textbf{0.8582} & \multirow{2}*{0.8657}\\
    ~ & ~ & \footnotesize{$[0.7978, 0.9003]$} & \footnotesize{$[0.8263, 0.9322]$} & \footnotesize{$[0.8043, 0.9127]$}&~\\
    \hline
  \end{tabular}}
\end{table}

Table \ref{tab1} summarizes the results. 
The first row presents the baseline model without modifications, 
where spectral and spatial features are fully extracted and combined. 
The second row adds the SE-block, while the third introduces the virtual class. 
The final row represents our full proposed method. 
Incorporating channel-wise attention increased the average AUC from $0.7652$ to $0.8153$. 
The virtual class injection boosted the baseline averaged AUC to $0.8239$. 
Our complete method achieved the highest AUC of $0.8657$, 
confirming the complementary effectiveness of both components in enhancing model performance.
\subsection{Impact of Single Energy-Level Inputs and Hidden Layer Optimization}
In the second ablation study, we assessed model performance using a single energy level as input. 
Since the SE-block requires multi-level input, 
only the effect of virtual class injection was evaluated. 
AUC results showed minimal variation across the eleven energy levels ($0.77\pm0.0015$), 
underscoring the diagnostic limitations of single-energy inputs. 
Hence, we selected 40 keV as the representative level to demonstrate the impact of virtual class. 
Without virtual class injection, the model achieved an average AUC of $0.77$ across folds, 
which improved to $0.83$ with the injection. 
Although the second fold (AUC $0.81$) performed relatively well with single-energy input, 
it still lagged behind the full model using all energy levels. 
These results highlight the value of the SE-block in leveraging cross-level correlations 
and further validate the effectiveness of the combined approach.

In addition, prior to inject the virtual class, 
we optimized the parameter in the channel-wise attention mechanism—specifically, 
the number of output nodes in the fully connected layer following the global average pooling (GAP) in the SE-block. 
Due to the inherent squeezing operation, this number was constrained to be $\leq11$ (the number of channels). 
We conducted eleven experiments, each with a different node count, 
and selected $4$ as the optimal value based on the highest average AUC.

\subsection{Comparative Analysis with Established CNN Architectures}\label{comparison}
In clinical practice, diagnosing metastasis in breast cancer patients typically follows a two-step process: 
first, identifying whether the case is metastatic or non-metastatic, 
and second, categorizing metastatic cases as either low-burden or heavy-burden. 
In contrast, CNN-based methods can directly distinguish among all three classes. 
To further validate the effectiveness of our proposed CNN framework, 
we compared its performance with several widely used and well-established CNN architectures 
commonly applied in computer vision image recognition tasks.

\begin{table}[htbp!]\label{tab2}
  \centering
  \caption{Comparison of test set AUC, parameter count, and average inference time across three folds.}
  \resizebox{\linewidth}{!}{
  \begin{tabular}{l|c|c|c|c|c|c}
      \hline
      \multirow{2}*{Method}&Parameter&Time&\multicolumn{3}{c|}{Test AUC - BCa 95\% CI}&Averaged\\
      \cline{4-6}
      ~ & (million) &(seconds)&  Fold-1 & Fold-2 & Fold-3 & AUC\\
      \hline
      \multirow{2}*{VGG16} & \multirow{2}*{50.319} & \multirow{2}*{$\approx1.390$} & 0.6640 & 0.6553 & 0.6992 & \multirow{2}*{0.6728}\\
      ~ & ~ & ~ & \footnotesize{[0.5470, 0.7581]} & \footnotesize{[0.5542, 0.7520]} & \footnotesize{[0.5935, 0.7939]}&~\\
      \hline
      \multirow{2}*{VGG19} & \multirow{2}*{55.701} & \multirow{2}*{$\approx1.126$} & 0.6506 & 0.6387 & 0.7040 & \multirow{2}*{0.6644}\\
      ~ & ~ & ~ & \footnotesize{[0.5389, 0.7519]} & \footnotesize{[0.5305, 0.7471]} & \footnotesize{[0.5984, 0.8002]}&~\\
      \hline
      \multirow{2}*{ResNet} & \multirow{2}*{23.619} & \multirow{2}*{$\approx5.695$} & 0.7585 & 0.6833 & 0.7180 & \multirow{2}*{0.7199}\\
      ~ & ~ & ~ & \footnotesize{[0.6484, 0.8472]} & \footnotesize{[0.5814, 0.7773]} & \footnotesize{[0.6209, 0.8202]}&~\\
      \hline
      \multirow{2}*{DenseNet} & \multirow{2}*{7.066} & \multirow{2}*{$\approx12.595$} & 0.7784 & 0.7091 & 0.6763 & \multirow{2}*{0.7213}\\
      ~ & ~ & ~ & \footnotesize{[0.6658, 0.8630]} & \footnotesize{[0.6062, 0.8047]} & \footnotesize{[0.5768, 0.7634]}\\
      \hline
      \multirow{2}*{\textbf{Proposed}} & \multirow{2}*{\textbf{0.006}} & \multirow{2}*{$\approx\textbf{0.560}$} & \textbf{0.8528} & \textbf{0.8861} & \textbf{0.8582} & \multirow{2}*{0.8657}\\
      ~ & ~ & ~ & \footnotesize{$[0.7978, 0.9003]$} & \footnotesize{$[0.8263, 0.9322]$} & \footnotesize{$[0.8043, 0.9127]$}&~\\
      \hline
  \end{tabular}}
\end{table}
Table \ref{tab2} compares the performance of our proposed method with several established CNN architectures, 
including VGG \cite{simonyan2014very}, ResNet \cite{he2016deep}, and DenseNet \cite{huang2017densely}. 
The final row highlights our method, 
which achieved the highest average AUC of $0.8657$ across three folds, 
demonstrating its superior effectiveness. 
Notably, it also required significantly fewer parameters (0.006 million) 
and achieved a lower average inference time (0.560 seconds), highlighting its efficiency.

\subsection{ROC Curve Analysis and Statistical Validation}
To provide a more detailed analysis, we compared the ROC curves of the evaluated methods. 
The ROC curve reflects the classification performance of the selected model, 
with a shorter distance to the top-left corner $(0, 1)$ indicating better classification ability. 
The x-axis represents the False Positive Rate (FPR), while the y-axis denotes the True Positive Rate (TPR).

\begin{figure}[htbp]
  \centering
  \subfigure[Ablation study of fold 1]{
    \includegraphics[width=0.34\linewidth]{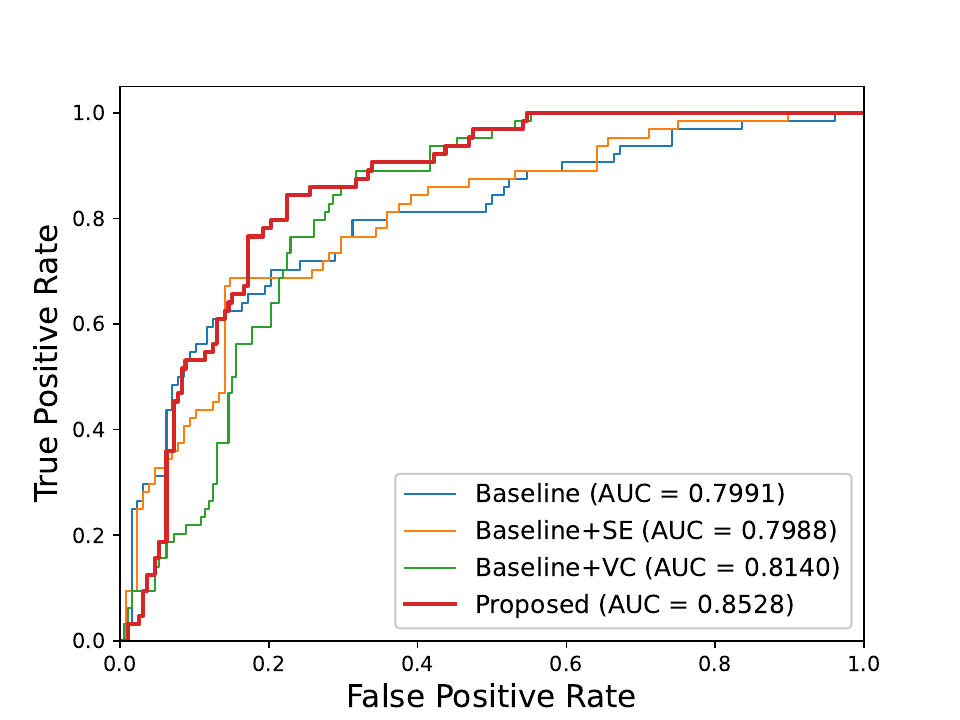}
  }\hspace{-8mm}
  \subfigure[Ablation study of fold 2]{
    \includegraphics[width=0.34\linewidth]{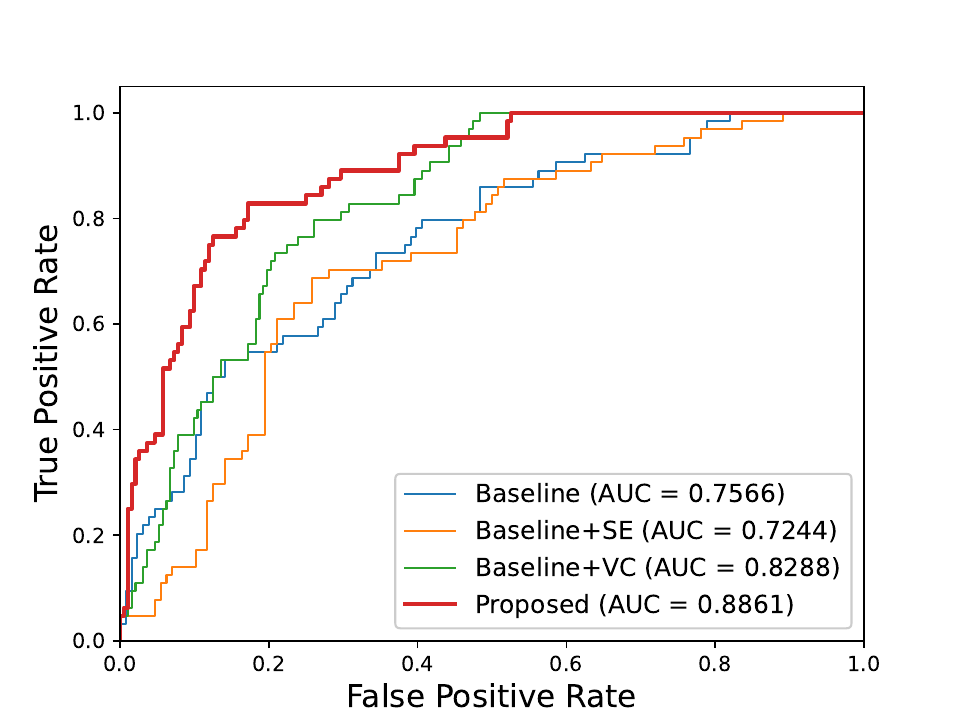}
  }\hspace{-8mm}
  \subfigure[Ablation study of fold 3]{
    \includegraphics[width=0.34\linewidth]{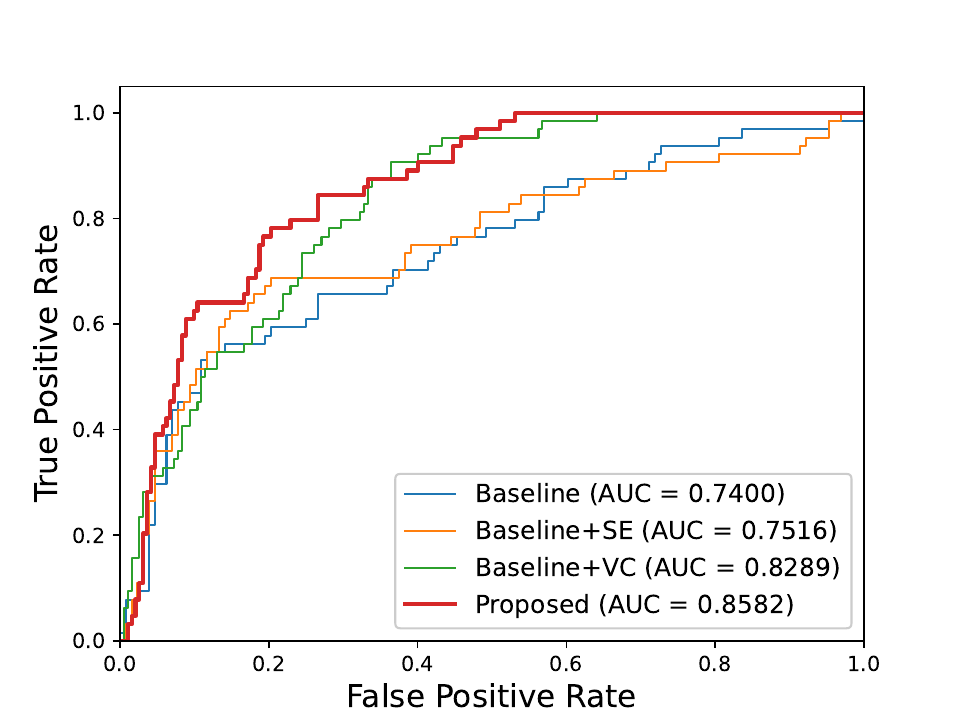}
  }
  % \quad
  \subfigure[Methods comparison of fold 1]{
    \includegraphics[width=0.34\linewidth]{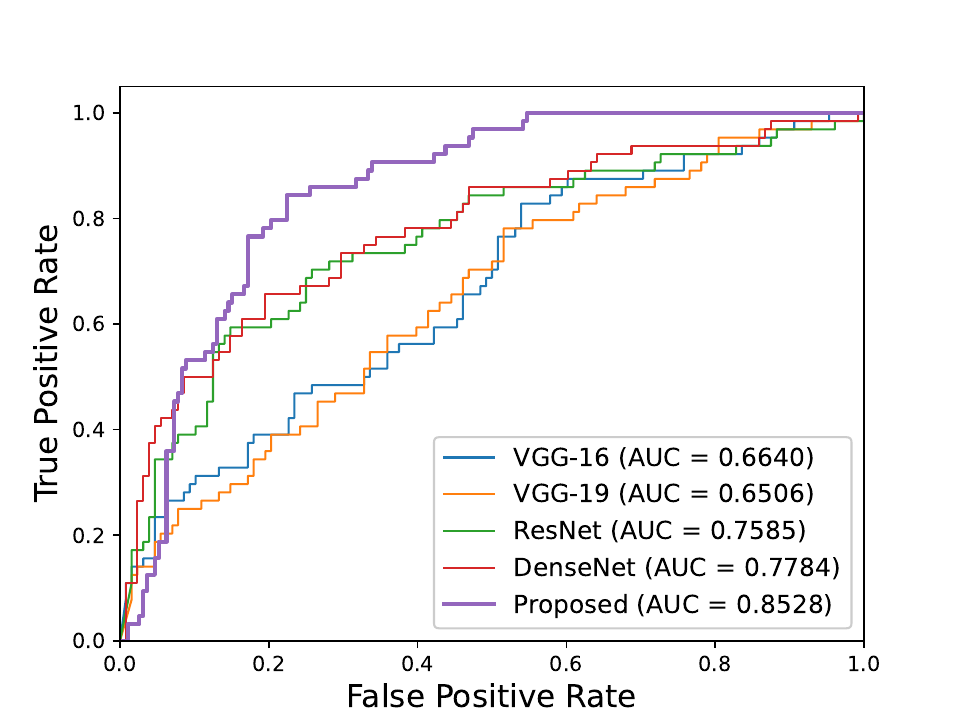}
  }\hspace{-8mm}
  \subfigure[Methods comparison of fold 2]{
    \includegraphics[width=0.34\linewidth]{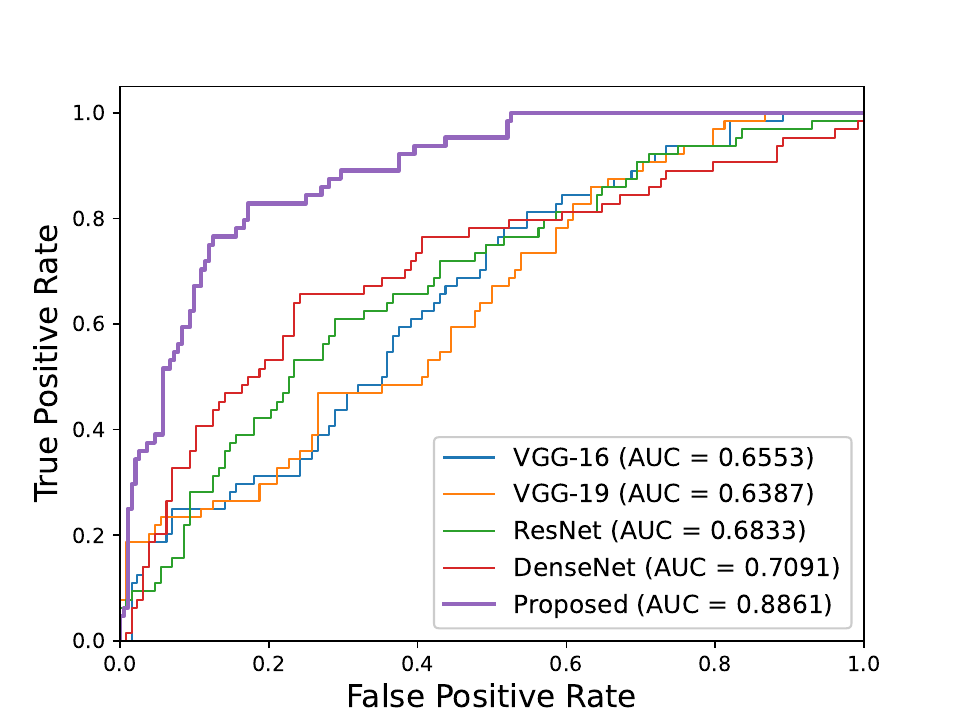}
  }\hspace{-8mm}
  \subfigure[Methods comparison of fold 3]{
    \includegraphics[width=0.34\linewidth]{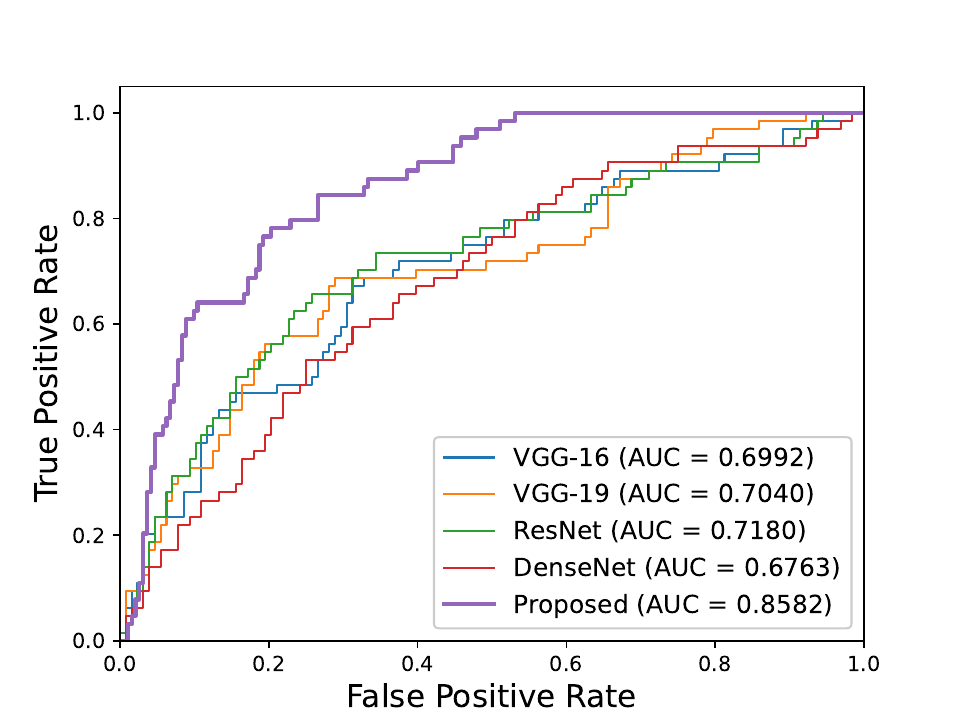}
  }
  \caption{Subfigures (a)-(c) show the ROC curves for the three folds in the ablation study, 
  while subfigures (d)-(f) compare the ROC curves of our proposed method with other CNN-based approaches.}
  \label{curve}
\end{figure}
Fig. \ref{curve} (a), (b), and (c) 
illustrate the ROC curves on the test sets for folds 1, 2, and 3, respectively, 
comparing the baseline model (blue line), 
the baseline with the SE-block (orange line), 
the baseline with virtual class injection (green line), 
and our proposed model (red line). 
Additionally, Fig. \ref{curve} (d), (e), and (f) 
present comparisons between our proposed method and other CNN-based approaches described in Subsection \ref{comparison}. 
In these figures, our method is shown in purple, while VGG-16, VGG-19, ResNet-50, 
and DenseNet-121 are represented by the blue, orange, green, and red lines, respectively.

In the ROC curves from our ablation study across the three folds (i.e., Fig. \ref{curve} (a), (b), and (c)), 
all models performed above the diagonal line (AUC $=0.5$), indicating effective classification. 
Both the baseline model with virtual class injection and our proposed method 
demonstrated strong performance (AUC $\geq0.8$). 
Although both the baseline model with the virtual class and our proposed method achieved high AUC scores, 
their ROC curves intersected—indicating that, in some regions, the baseline model yielded higher TPR at the same FPR. 
To further investigate, we applied DeLong's test \cite{delong1988comparing} 
on the AUC values across three folds and analyzed the class-wise p-values. 
In fold 1, our method outperformed the baseline for classes $N_0$, $N_{+(1-2)}$, and $N_{+(\geq3)}$, 
with AUCs of $0.8175$, $0.7109$, and $0.7153$, respectively,
compared to $0.7986$, $0.6733$, and $0.6175$ ($|z|>1.96$, $p<0.05$). 
In fold 2, both models performed comparably on $N_0$ and $N_{+(1-2)}$ ($|z|>1.96$, $p<0.05$), 
while results for $N_{+(\geq3)}$ were not statistically significant 
($|z|<1.96$, $p=0.053$, corrected $p=0.1613$ after Bonferroni correction). 
In fold 3, our method again outperformed the baseline across all classes, 
achieving AUCs of $0.7718$, $0.7782$, and $0.7581$, 
versus $0.6756$, $0.6308$, and $0.6449$ ($|z|>1.96$, $p<0.05$).

Subfigures (d)-(f) compare our proposed method with other well-known CNN-based models across the three folds. 
Among all methods, 
only our proposed approach consistently achieved strong performance, with AUC values $\geq0.8$, 
demonstrating the effectiveness of our framework. 
Notably, in folds 2 and 3, our proposed model consistently achieved higher TPR 
at the same FPR compared to the other methods. 

\section{Discussion}
% \subsection{Clinical Implications of Multi-Class Metastatic Burden Classification}

Our previous work demonstrated the effectiveness of a CNN-based multi-class classification framework 
for assessing sentinel axillary lymph node metastasis in breast cancer using 
a specialized in-house dual-energy computed tomography (DECT) dataset. 
This pathology-confirmed dataset enables categorization into three groups: 
no metastasis ($N_0$), low metastatic burdern ($N_{+(1-2)}$), and heavy metastatic burden ($N_{+(\geq3)}$). 
Accurate identification of $N_{+(1-2)}$ cases may help patients avoid unnecessary axillary lymph node dissection, 
thereby improving clinical outcomes. 
Unlike existing approaches that rely on multiple data modalities—often impractical for small datasets 
or burdensome for radiologists—our method leverages DECT's intrinsic spectral and spatial information. 
DECT enables detailed tissue characterization, making it ideal for this task. 
Building on our previous shallow binary classification model, 
which used $1\times1$ and $3\times3$ convolutions 
to decouple spectral and spatial features from $176\times176\times11$ tensors, 
we expanded the architecture to support multi-class classification, 
effectively addressing the first two challenges.
\begin{figure}[htbp]\label{fig2}
  \centering
  \includegraphics[width=0.98\textwidth]{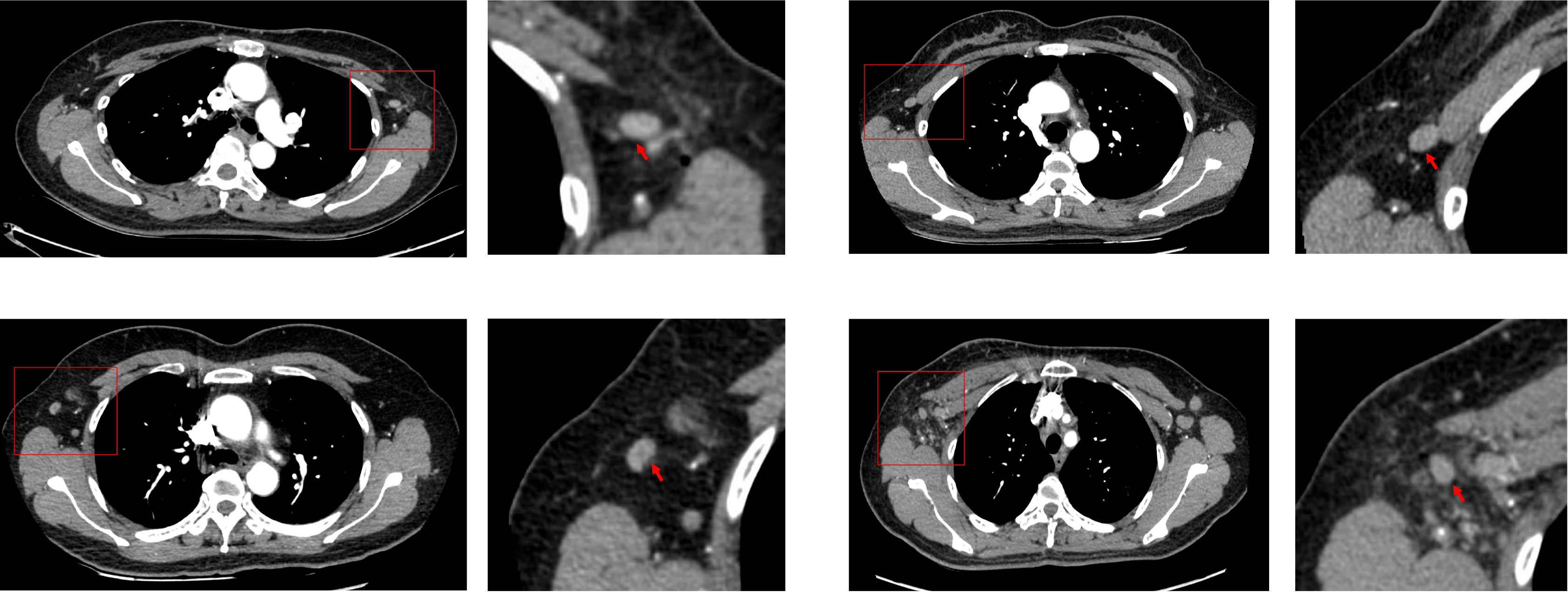}
  \caption
  {The first row shows patients with no metastasis $N_0$ and low-burden metastasis $N_{+(1-2)}$, 
  while the second row presents cases with heavy-burden metastasis $N_{+(\geq3)}$. 
  Sentinel axillary lymph nodes, highlighted by red bounding box, 
  illustrate the morphological similarities across different classes.}
\end{figure}
% \subsection{Integration of Features and Representation Refinement}

After addressing the initial two challenges, 
we further leveraged the multi-energy nature of DECT images (Fig. \ref{fig1}) 
and the morphological similarities among different lymph node classes (Fig. \ref{fig2}). 
To tackle these complexities, we focused on enhancing both the feature and representation spaces. 
We first introduced a feature space-squeezing strategy, illustrated through flowcharts and formulas, 
which compresses information across 11 energy levels (40-140 keV, 10 keV stride). 
This strategy encodes and re-excites feature maps using learnable coefficients from a shallow embedding layer, 
improving representational capacity by refining spectral integration. 
To address the final challenge, 
we incorporated a virtual class mechanism by injecting a corresponding fake label into the representation space. 
Given the visual similarity across classes, 
this approach sharpens inter-class boundaries and compacts intra-class variations, enhancing class discrimination. 
It also incorporates a theoretical foundation through the formulation of a virtual softmax loss function. 
Together, these innovations strengthen the model's theoretical grounding and practical utility, 
supporting reliable clinical application in lymph node metastasis diagnosis.

Experimental results confirmed the effectiveness of our backbone design and underscored the performance gains achieved 
through the integration of the proposed space-squeezing strategies. 
Ablation studies, numerical comparisons, 
and ROC curve analyses demonstrated the efficiency and robustness of the overall framework. 
These findings strongly support the potential of our model as a reliable, non-invasive, 
and effective computer-aided diagnostic tool for breast cancer patients in clinical settings.
% \subsection{Limitations and Future Directions}

The regions of interest (RoI) in our study were limited to sentinel axillary lymph nodes, 
providing only localized information for breast cancer patients. 
Future research should broaden the scope to include additional lymph nodes, 
enabling a more comprehensive assessment of nodal status by fully accounting for all relevant nodes. 
Furthermore, developing more sophisticated models with additional perceptron layers or modules will be essential 
for enhancing the generalizability and robustness of the predictive framework. 

\section{Conclusions}
This study introduces a DECT-based space-squeeze method that advances multi-class 
classification of metastatic lymph nodes in breast cancer. 
By compressing spectral-spatial features through 
channel-wise attention and refining representation space via virtual class injection, 
our approach achieves superior performance (averaged-AUC: 86.57\%) 
in distinguishing $N_0$, $N_{+(1-2)}$, and $N_{+(\geq3)}$ categories. 
The method's ability to reduce diagnostic ambiguity underscores 
its clinical potential for guiding personalized treatment strategies, 
particularly in avoiding unnecessary axillary dissections for low-burden cases. 
Future work will expand the model's scope to include auxiliary lymph nodes and 
integrate advanced architectures for broader applicability. 
This framework establishes a robust baseline for leveraging DECT's unique

% \clearpage

\section*{Acknowledgements}
% This work was supported by National Key Research and Development Program of China [No. 2023YFE0204300], 
% National Natural Science Foundation of China 
% [No. 82441027, 62371476, 82371917, 82102130], 
% Guangzhou Science and Technology Bureau [No. 2023B03J1237], 
% R\&D Program of Pazhou Lab (HuangPu) [No. 2023K0606], 
% Guangdong Basic and Applied Basic Research Foundation [No. 2023A1515011305], 
% Guangzhou Basic and Applied Basic Research Foundation [No. 2023A04J2112, 202201011268], 
% Health Research Major Projects of Hunan Health Commission [No. W20241010], 
% Guangdong Province Key Laboratory of Computational Science at the Sun Yat-sen University [No. 2020B1212060032].
\section*{References}
\addcontentsline{toc}{section}{\numberline{}References}
\vspace*{-20mm}
\bibliography{./example}    
% \begin{thebibliography}{10}

%   \end{thebibliography}
\bibliographystyle{./medphy.bst}    
\end{document}